\documentclass[12pt,a4paper]{article}
\usepackage[]{amsmath,amssymb}
\usepackage{graphicx}
\usepackage{geometry}
\usepackage{amssymb,epsfig,subfigure}
\usepackage{graphics,epsfig}
\usepackage{graphicx, color}
\usepackage{geometry}
\usepackage{diagbox}
\usepackage{float}
\usepackage{xcolor}
\usepackage{bbold}
\usepackage[latin1]{inputenc}
\usepackage[font=footnotesize]{caption}
\usepackage[T1]{fontenc}
\usepackage{amsmath}
\usepackage{amsfonts}
\usepackage{amssymb}
\usepackage{latexsym}
\usepackage[english]{babel}
\usepackage{authblk}
\newcommand{\be}{\begin{equation}}
\newcommand{\ee}{\end{equation}}
\newcommand{\beq}{\begin{equation}}
\newcommand{\eeq}{\end{equation}}
\newcommand{\bea}{\begin{eqnarray}}
\newcommand{\eea}{\end{eqnarray}}

\def\be{\begin{equation}}
\def\ee{\end{equation}}
\def\ba{\begin{eqnarray}}
\def\ea{\end{eqnarray}}

\definecolor{princetonorange}{rgb}{1.0, 0.56, 0.0}
\definecolor{WildStrawberry}{rgb}{1.0, 0.26, 0.64}
\definecolor{rossocorsa}{rgb}{0.83, 0.0, 0.0}
\definecolor{navyblue}{rgb}{0.0, 0.0, 0.5}
\usepackage{hyperref}

\hypersetup{
    colorlinks,
    citecolor=rossocorsa,
    filecolor=navyblue,
    linkcolor=navyblue,
    urlcolor=navyblue
}

\begin{document}
\title{Modular Hamiltonian of the scalar in the semi infinite line: dimensional reduction for spherically symmetric regions}

\author{Marina Huerta\footnote{e-mail: marina.huerta@cab.cnea.gov.ar} }

\author{Guido van der Velde\footnote{e-mail: guido.vandervelde@ib.edu.ar }}
\affil{Centro At\'omico Bariloche, 8400-S.C. de Bariloche, R\'{\i}o Negro, Argentina}

\date{}

\maketitle
\begin{abstract}
We focus our attention on the one dimensional scalar theories that result from dimensionally reducing the free scalar field theory in arbitrary $d$ dimensions. As is well known, after integrating out the angular coordinates, the free scalar theory can be expressed as an infinite sum of theories living in the semi-infinite line, labeled by the angular modes $\{\ell,\vec{m}\}$. 
 We show that their modular Hamiltonian in an interval attached to the origin is, in turn, the one obtained from the dimensional reduction of the modular Hamiltonian of the conformal parent theory in a sphere. Remarkably, this is a local expression in the energy density, as happens in the conformal case, although the resulting one-dimensional theories are clearly not conformal. We support this result by analyzing the symmetries of these theories, which turn out to be a portion of the original conformal group, and proving that the reduced modular Hamiltonian is in fact the operator generating the modular flow in the interval. By studying the spectrum of these modular Hamiltonians, we also provide an analytic expression for the associated entanglement entropy. Finally, extending the radial regularization scheme originally introduced by Srednicki, we sum over the angular modes to successfully recover the conformal anomaly in the entropy logarithmic coefficient in even dimensions, as well as the universal constant $F$ term in $d=3$.

\end{abstract}


\section{Introduction: Modular flow and modular Hamiltonian} \label{intro}

The successful application of information theory tools to quantum field theory (QFT) along the last decades, has given place to the solid current consensus that these tools must be definitively incorporated into the usual QFT machinery. In this context, the study of quantities related to different information measures for quantum field theories gains relevance and with them, the study of states reduced to a region. These states are described by reduced (local) density matrices that live in the core of the definition of all the information measures referenced to spatial regions $R$. From the quantum field algebraic perspective \cite{Haag}, each region $R$ is attached to the algebra of the degrees of freedom localized in $R$. The reduced state to a local algebra of operators in a region can be expressed, in presence of a cutoff, as a density matrix 
\be
\rho= \frac{e^{-K}}{\textrm{tr}e^{-K}}\,, \label{termal}
\ee
where the exponent $K$ is the modular Hamiltonian operator. This convenient way of
encoding the reduced state admits an interesting interpretation of the entanglement entropy as the thermodynamic entropy of a system in equilibrium at temperature $1$, but with respect to the modular Hamiltonian $K$. Moreover, there is a time notion associated to the state through the modular Hamiltonian, whose evolution is implemented by the unitary operator in the algebra
\be
U(\tau)=\rho^{i \tau}\sim e^{-i \tau K}\,. \label{modflow}
\ee      
The induced evolution of operators $O(\tau)= U(\tau) O U(-\tau)$ is called the modular flow. This is a purely quantum transformation, which becomes trivial in the classical limit. 
 
Historically, the earliest recognition of the structural importance of modular flows can be found in the algebraic formulation of QFT \cite{brunetti1993modular,Hislop:1981uh} and more recently, in the framework of the study of different information measures and statistical properties of reduced states in QFT \cite{Casini:2017roe,Arias:2017dda, ohya2004quantum}. The modular Hamiltonian is a fundamental constitutive part of the relative entropy and plays an essential role in the entropy bounds formulations and proof of several energy conditions \cite{Casini:2008cr,Wall:2011hj,Blanco:2013lea,Bousso:2014uxa,Bousso:2015wca,Hollands:2018wzi}.
Besides, profiting that entanglement and relative entropy have well established geometric duals for holographic QFT \cite{Ryu:2006bv,Ryu:2006ef, Jafferis:2015del}, modular Hamiltonians have also been used to clarify localization properties of degrees of freedom in quantum gravity \cite{Blanco:2013joa,Jafferis:2014lza,Faulkner:2017vdd}.

Currently, our knowledge of the explicit form of modular Hamiltonians reduces mostly to some examples where the modular flow is local, and it is primarily determined by spacetime symmetries. 

This is the case for the Rindler wedge $x^1>\vert t \vert$ in Minkowski space and any QFT. Choosing the causal region to be the half spatial plane $x^1>0$ and $t=0$ then, the rotational symmetry of the euclidean theory allows us to express the reduced density matrix corresponding to the vacuum state in terms of the energy density $T_{00}$
\be
\rho = k\, e^{-2\pi \int_{x^1>0}d^{d-1}x\,x^1 T_{00}(x)}\,.
\label{rindler}
\ee

The above expression manifestly reveals a non trivial connection between entanglement in vacuum and energy density.
Moreover, in equation (\ref{rindler}), the exponent corresponds to the modular Hamiltonian for half space which results to be an integral of a local operator. $K$ is in fact $2\pi$ times the generator of boosts restricted to act only on the right Rindler wedge
\be
K=-2\pi \int_{x^1>0}d^{d-1}x\,x^1 T_{00}(x)\,.
\ee
The modular flow $\rho^{i\tau}$ moves operators locally following the orbits of the one parameter group of boost transformations.
On the other hand, it is interesting to note that from equation (\ref{rindler}), the vacuum state in half space corresponds to a thermal state of inverse temperature $2\pi$ with respect to the boost operator. 
This is directly connected to the Unruh's effect \cite{unruh1976notes} according to which accelerated observers see the vacuum as a thermally excited state.
For an observer following a trajectory given by a boost orbit, the state looks like a thermal state with respect to the proper time $\tilde{\tau}$. For these trajectories, the proper time and the boost parameter $s$ are proportional $s=a \tilde{\tau}$ with $a$ the proper acceleration of the observer, constant along boost orbits. In turn, this implies there is a relation $K= \tilde{H}/a$ between the boost operator and the proper time Hamiltonian $\tilde{H}$ of the accelerated observer. For such an observer there is a thermal bath at (proper time) temperature $T=\frac{2\pi}{a}$.

The other very well known example where symmetries again facilitate the derivation of the exact modular Hamiltonian is the case of conformal field theories (CFT)  for spheres in any dimensions.
For a CFT, Poincare symmetries are enlarged to the conformal group. These theories are characterized by having a traceless, symmetric and conserved stress tensor. This enlarges the number of conserved currents related to space-time symmetries which in general can be written as
\be  
j_\mu= a^\nu \,T_{\nu\mu} +  b^{\alpha \nu}\,x_\alpha\, T_{\nu\mu} + c\, x^\nu\, T_{\nu\mu} + d_\alpha\, (x^2 g^{\alpha\nu}-2\, x^\alpha x^\nu)\, T_{\nu\mu}\,.\label{das} 
\ee
The corresponding conserved charges depend on parameters $a^\mu$, determining translations, the antisymmetric $b^{\mu\nu}$, giving Lorentz transformations, $c$, related to dilatations, and $d^\mu$, for the so called special conformal transformations.   

Since there is a conformal transformation that maps the Rindler wedge to causal regions with spherical boundary, and the same transformation leaves the vacuum invariant for a CFT, then, the modular Hamiltonian is just the transformed Rindler  modular Hamiltonian. It is easy to  get
\be
K=2\pi \int_{|\vec{x}|<R} d^{d-1}x\, \frac{R^2-r^2}{2R}\, T_{00}(\vec{x})\,.\label{modesf}
\ee    
In this example, $K$ is again local and proportional to $T_{00}$, with a proportionality weight function $\beta(r)\equiv\frac{R^2-r^2}{2R}$.

Except for the two examples discussed above, the vacuum of a QFT in the Rindler wedge and the vacuum of a CFT in the sphere, there are only some other few known modular Hamiltonians, either local or not. The local ones in general are derived profiting from symmetry transformations that leave the state invariant. This is for example the case of the modular Hamiltonian for CFTs in $1+1$ dimensions in presence of a global
or local quench \cite{Cardy:2016fqc,Tonni:2017jom,DiGiulio:2019lpb,DiGiulio:2019cxv,Eisler:2020lyn,Mintchev:2020jhc,Mintchev:2020uom}.  
However, on general grounds, from the point of view of quantum information we do not expect locality to hold. In general, $K$ will be given by a non local
and non linear combination of the field operators at different positions inside the region.

An example of a non local modular
Hamiltonian which has been explicitly computed is the one for the vacuum state of the free massless fermion in $d = 2$ for several disjoint intervals
\cite{Casini:2009vk,Longo:2009mn,Wong:2018svs}. In this case $K$ has a local term proportional to the energy density and an additional non local part given by a quadratic expression in the fermion field that connects in a very particular way points located in different intervals.

In this paper we calculate the modular Hamiltonian for the vacuum state of non conformal $(1+1)$ dimensional theories in the interval $(0,R)$. These theories are defined in the semi infinite line, and result from the dimensional reduction of the $d$ dimensional free massless scalar.
Our strategy is to calculate the modular Hamiltonian of the reduced system by profiting of the known modular Hamiltonian of CFTs in spheres in any dimension. 

\begin{figure}[t]
\begin{center}  
\includegraphics[width=0.50\textwidth]{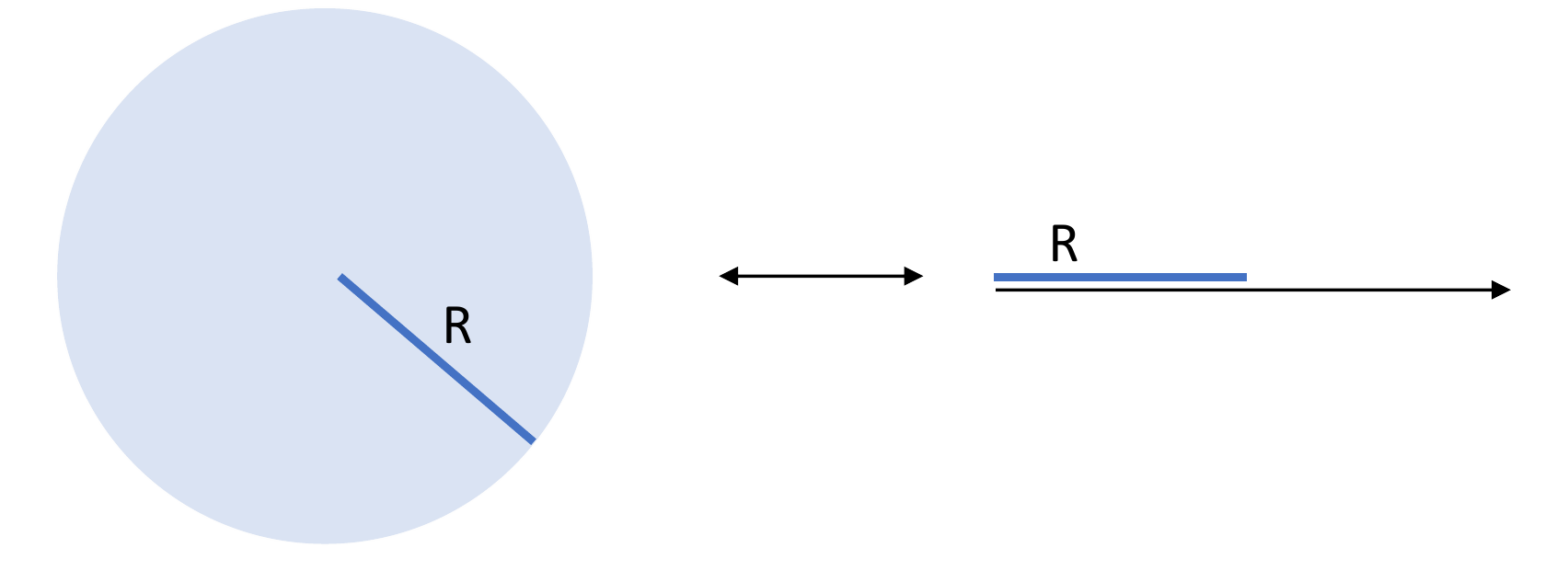}
\captionsetup{width=0.9\textwidth}
\caption{The sphere of radius $R$ corresponds to intervals of length $R$ with one edge in the origin in the radial semi infinite line. }
\label{sphere_interval}
\end{center}  
\end{figure}

The free massless scalar in $d$ space time dimensions can be dimensionally reduced to a sum of one dimensional theories, one for each angular mode. Since the reduction is obtained by integrating over the angular coordinates, these systems live in the semi infinite line. From the algebraic point of view, this is convenient when studying algebras assigned to spherical regions to calculate, for example, the entanglement entropy. In these coordinates, the local algebra assigned to the region can be easily written in terms of fields $\phi(r,\Omega)$ with nice localization properties. For example, points in the semi infinite line correspond to shells in the original space and intervals connected to the origin, to $d$-spheres (see figure \ref{sphere_interval}).
Concretely, in the radial coordinate, the canonical Hamiltonian for the massless free scalar decomposes as a sum over angular modes $H_{\ell \vec{m}}$
\be
H=\sum_{\ell \vec{m}}H_{\ell \vec{m}}\,.
\ee
with $(\ell \vec{m})$ the angular mode label.
In fact, there is a family of one dimensional Hamiltonian $H_{\ell \vec{m}}$ for each dimension.
In turn, the same decomposition occurs for the modular Hamiltonian (\ref{modesf})
\be
K=\sum_{\ell \vec{m}}K_{\ell \vec{m}}\,.
\ee
Taking into account that the vacuum state for a system composed by independent subsystems is a product of density matrices, here $\rho=\otimes\rho_{\ell \vec{m}}$, then it is immediate to identify the modular Hamiltonian mode $K_{\ell \vec{m}}$ with the modular Hamiltonian of the one dimensional reduced system $H_{\ell \vec{m}}$. The Hamiltonian $H_{\ell \vec{m}}$ does not correspond to a conformal relativistic theory due to an extra quadratic term proportional to $1/r^2$, whose proportionality constant depends on the dimension of the original problem and the angular mode $\ell$\footnote{For a different context in which a free scalar, albeit conformal, is obtained from dimensional reduction, see \cite{Edery:2009vr}.}.
Surprisingly, we find that $K_{\ell\vec{m}}$ is still local and proportional to the energy density $T_{00}$\footnote{Since translational invariance is lost, there is no conserved energy momentum tensor. The notation for the energy density is just a matter of convention.}, with the same weight function $\beta(r)$ that characterizes the modular Hamiltonian for CFTs in spheres.
Our analytic results coincide with the suggested continuum limit of the entanglement Hamiltonian of blocks of consecutive sites in massless harmonic chains, recently studied in \cite{Javerzat:2021hxt}.

This article is organized as follows. In section \ref{sec:Spherical coordinates} we explicitly carry out the dimensional reduction. We write the scalar field in a basis of hyper-spherical harmonics, and after integrating out the angular coordinates we are left with a Hamiltonian for the reduced systems $H_{\ell \vec{m}}$ of the form
\be
H_{\ell \vec{m}}=\frac{1}{2}\int  dr \left[\widetilde{\pi}_{\ell\vec{m}}^2 +(\partial_r \widetilde{\phi}_{\ell\vec{m}})^2+\frac{\mu_d(\ell)}{r^2} \widetilde{\phi}_{\ell\vec{m}} ^2 \right],
\label{h1d}
\ee
with
\be
\mu_d(\ell)=\frac{(d-4)(d-2)}{4}+\ell(\ell+d-3). 
\ee
In section \ref{sec: mH} the same procedure is followed to find the modular Hamiltonian
\be\label{Kintro}
K_{\ell, \vec{m}}=2\pi \int_{|\vec{x}|<R} dr\, \frac{R^2-r^2}{2R}\, T_{00}^{\ell, \vec{m}}(\vec{x})\,.
\ee 
In some way, the reduced theory, manifestly invariant under dilatations but non conformal, keeps the \textit{memory} of the conformal symmetry of the parent $d$-dimensional theory \cite{Jackiw:2011vz,Chamon:2011xk}, with the same local modular Hamiltonian as that representing the vacuum of a CFT in a sphere. We delve into this in section \ref{sec:symmetries}, where we show that the reduced theories preserve an $SL(2,\mathbb{R})$ symmetry, and that the modular transformation belongs to this subgroup.  The modular Hamiltonian (\ref{Kintro}) written as a Noether charge can be correctly interpreted as the local operator implementing the modular flow.

In section \ref{sec:entropy} we solve the spectrum of the modular Hamiltonian (\ref{Kintro}) and compute the entanglement entropy in a segment connected to the origin. We find the analytic expression
\begin{equation}
S(\ell,d)=\frac{1}{6}\log{\frac{R}{\epsilon}}-\frac{i \pi}{2}\int_0^{\infty} ds \frac{s}{\sinh^2(\pi s)}\log{\left(\frac{4^{ i s}\Gamma\left[i s\right]\Gamma\left[-1+d/2+\ell-i s\right]}{\Gamma\left[-i s\right]\Gamma\left[-1+d/2+\ell+i s\right]}\right)},
\end{equation}
which is logarithmically divergent, with coefficient $1/6$ as expected for $(1+1)$ theories, and has a constant term that depends both on the mode $\ell$ and the space time dimensions $d$ of the original theory. Although the above integral cannot in general be solved analytically, we make some useful approximations to extract relevant information out of it. Moreover, by summing over $\ell$ we are able to recover the conformal anomaly in the logarithmic coefficient for the free scalar field in even dimensions, as well as the constant universal $F$ term in $d=3$. In doing the sum over the angular modes $\ell$, we introduce a novel regularization implemented by a damping exponential $\exp[-\ell\epsilon/R]$, with the same cutoff $\epsilon$ that regularizes the radial coordinate $r$. This procedure generalizes the radial regularization scheme introduced by Srednicki in \cite{Srednicki:1993im}, where it is explicitly stated that for $d \geqslant 4$  regularization by a radial lattice turns out to be insufficient and the sum over partial waves does not converge.
We end the discussion with some concluding remarks.

\section{Spherical coordinates}\label{sec:Spherical coordinates}
The free scalar action in spherical coordinates reads
\begin{equation}
S=\frac{1}{2}\int dt dr r^{d-2}d\Omega \left[-(\partial_0 \phi)^2+ (\partial_r \phi)^2- \frac{\phi}{r^2}\Delta_{S^{d-2}}\phi\right].
\end{equation}
With the aim of reducing the above to a single integral in the radial direction, we Fourier transform the scalar field in the angular coordinates, using the real hyper-spherical harmonics as basis functions,
\begin{equation}
\phi(\vec{r})=\sum_{\ell m_1 ... m_{d-3}}\phi_{\ell m_1 ... m_{d-3}}(r) Y_{\ell}^{m_1 ... m_{d-3}}(\hat{r}),
\end{equation}
with
\begin{equation}
\Delta_{S^{d-2}}Y_{\ell}^{m_1 ...m_{d-3}}(\hat{r})=-\ell(\ell+d-3)Y_{\ell}^{m_1 ...m_{d-3}}(\hat{r}),
\end{equation}
\begin{equation}
\int_{S^{d-2}}d\Omega Y_{\ell}^{m_1 ... m_{d-3}}(\hat{r}){Y}_{\ell '}^{m_1 ' ... m_{d-3}'}(\hat{r})=\delta_{\ell\ell '}\delta_{m_1 m_1'}...\delta_{m_{d-3} m_{d-3}'}.
\end{equation}
After integrating the angular coordinates, we are left with
\begin{equation}
S=\frac{1}{2}\sum_{\ell \vec{m}}\int dt dr r^{d-2} \left[-(\partial_0 \phi_{\ell\vec{m}})^2 +(\partial_r \phi_{\ell\vec{m}})^2+\frac{\ell(\ell+d-3)}{r^2} \phi_{\ell\vec{m}}^2 \right].
\end{equation} 
However, the theory looks simpler when defined in terms of the rescaled field $\widetilde{\phi}_{\ell\vec{m}}=r^{\frac{d-2}{2}}\phi_{\ell\vec{m}}$, whose canonically conjugated momentum is $\widetilde{\pi}_{\ell\vec{m}}\equiv \partial_0 \widetilde{\phi}_{\ell\vec{m}}$,
\begin{equation}\label{S1}
S=\frac{1}{2}\sum_{\ell \vec{m}}\int dt dr \left[-(\partial_0 \widetilde{\phi}_{\ell\vec{m}})^2 +r^{d-2}\left(\partial_r \left(\frac{\widetilde{\phi}_{\ell\vec{m}}}{r^{\frac{d-2}{2}}}\right)\right)^2+\frac{\ell(\ell+d-3)}{r^2} \widetilde{\phi}_{\ell\vec{m}}^2 \right].
\end{equation}
Functional variation with respect to the field leads to the equation of motion. Nevertheless, in order for the variational problem to be well posed we should impose specific boundary conditions at $r=0$. In fact,
\begin{equation}\label{sol}
\begin{split}
\delta S= \sum_{\ell\vec{m}}&\left\{\int dt dr \left[\partial_0^2 \widetilde{\phi}_{\ell\vec{m}}-\frac{1}{r^{\frac{d-2}{2}}}\partial_r \left(r^{d-2} \partial_r\left(\frac{\widetilde{\phi}_{\ell\vec{m}}}{r^{\frac{d-2}{2}}}\right)\right)+\frac{\ell(\ell+d-3)}{r^2}\widetilde{\phi}_{\ell\vec{m}}\right]\delta {\widetilde{\phi}}_{\ell\vec{m}}\right.\\
& \left.+ \int dt\left[r^{\frac{d-2}{2}}\partial_r \left(\frac{\widetilde{\phi}_{\ell\vec{m}}}{r^{\frac{d-2}{2}}}\right)\delta{\widetilde{\phi}}_{\ell\vec{m}}\right]\Biggr\rvert_0^{\infty}\right\}\, ,
\end{split}
\end{equation}
which requires either $\delta \widetilde{\phi}_{\ell\vec{m}}(r=0,t)=0$ (Dirichlet boundary conditions) or $r^{\frac{d-2}{2}}\partial_r \left(\frac{\widetilde{\phi}_{\ell\vec{m}}}{r^{\frac{d-2}{2}}}\right)\rightarrow 0$ (analogous to the ordinary Neumann boundary conditions). In the following we will adopt the former.

The second term in (\ref{sol}) can be further simplified, which leads to the saddle point
\begin{equation}
\partial_0^2 \widetilde{\phi}_{\ell\vec{m}}-\partial_r^2 \widetilde{\phi}_{\ell\vec{m}}+\frac{\mu_d(\ell)}{r^2}\widetilde{\phi}_{\ell\vec{m}}=0,
\end{equation}
with 
\begin{equation}
\mu_d(\ell)=\frac{(d-4)(d-2)}{4}+\ell(\ell+d-3).
\end{equation}
This partial differential equation can be solved by separation of variables, and expressed in terms of the original field $\phi_{\ell\vec{m}}$, the radial eigenfunction problem is a Bessel equation, with solution $j_{\ell}(r)\equiv \frac{1}{r^{(d-3)/2}}J_{\ell+\frac{d-3}{2}}(k r)$. Therefore, the solution is
\begin{equation}\label{classical}
\widetilde{\phi}_{\ell\vec{m}}(t,r)=e^{\pm i k t}\sqrt{k r}J_{\ell+\frac{d-3}{2}}(k r),
\end{equation}
which means that $\widetilde{\phi}_{\ell\vec{m}}\sim r^{\ell +d/2-1}$ near $k r\sim 0$, in agreement with the boundary conditions.

Having stated that, it is also possible to rewrite the second term in (\ref{S1}) by getting rid of a boundary term\footnote{We would be able to ignore the boundary term provided $ \widetilde{\phi}_{\ell\vec{m}} ^2$ went to zero faster than $r$. This is at least satisfied by the classical configuration (\ref{classical}).}. More explicitly, 
\begin{equation}
S=\sum_{\ell \vec{m}}S_{\ell \vec{m}}
\end{equation}
where 
\begin{equation}\label{action1}
S_{\ell \vec{m}}=\frac{1}{2}\int dt dr \left[-(\partial_0 \widetilde{\phi}_{\ell\vec{m}})^2 +(\partial_r \widetilde{\phi}_{\ell\vec{m}})^2+\frac{\mu_d(\ell)}{r^2} \widetilde{\phi}_{\ell\vec{m}} ^2 \right]
\end{equation}
can be thought of as the action for a free scalar living in the half line, satisfying Dirichlet boundary conditions at the origin. Note that, unlike the theory we started with, this is not a CFT because of the last term.

The dimensional reduction of the free scalar Hamiltonian can be made following the same steps. But we can alternatively calculate the conserved charge due to time translations associated directly to the $1+1$ dimensional action (\ref{action1}), yielding 
\begin{equation}\label{Hscalar}
H=\frac{1}{2}\sum_{\ell \vec{m}}\int  dr \left[\widetilde{\pi}_{\ell\vec{m}}^2 +(\partial_r \widetilde{\phi}_{\ell\vec{m}})^2+\frac{\mu_d(\ell)}{r^2} \widetilde{\phi}_{\ell\vec{m}} ^2 \right]
\end{equation}
Once again we stress that $\widetilde{\pi}_{\ell\vec{m}}$ and $\widetilde{\phi}_{\ell\vec{m}}$ satisfy canonical commutation relations
\begin{equation}
\left[\widetilde{\phi}_{\ell\vec{m}}(r),\widetilde{\pi}_{\ell'\vec{m}'}(r')\right]=i \delta_{\ell,\ell '}\delta_{\vec{m},\vec{m} '}\delta(r-r').
\end{equation}
\section{The sphere modular Hamiltonian}\label{sec: mH}
On the other hand, since the free scalar field theory in $d$ spacetime dimensions is conformally invariant, when the whole system is in its ground state the modular Hamiltonian of a sphere is
\begin{equation}\label{KCFT}
K=\frac{1}{2}\int_{\vert x\vert<R} dx^{d-1} \left(\frac{R^2-r^2}{2 R}\right)T_{00}.
\end{equation}
However, although the stress tensor involved in this expression must be traceless, the canonical stress tensor of the free scalar field is
\begin{equation}
T_{\mu\nu}^{(c)}=\partial_{\mu}\phi\partial_{\nu}\phi-\frac{1}{2}\eta_{\mu\nu}(\partial\phi)^2,
\end{equation}
which has non vanishing trace $T_{\mu}^{\mu}=\left(1-d/2\right)(\partial\phi)^2=\frac{(1-d/2)}{2}\partial^2(\phi^2)$\footnote{This identity holds on-shell.}. Hence, it must be improved by adding a conserved symmetric tensor. A possible choice is
\begin{equation}
T_{\mu\nu}'=T_{\mu\nu}^{(c)}-\frac{(1-d/2)}{2(1-d)}(\partial_{\mu}\partial_{\nu}-\eta_{\mu\nu}\partial^2)\phi^2.
\end{equation}
Therefore,
\begin{equation}
K=\frac{1}{2}\int_{\vert x\vert<R} dx^{d-1} \left(\frac{R^2-r^2}{2 R}\right)\left[(\partial_0 \phi)^2+(\partial_i \phi)^2-\frac{(1-d/2)}{(1-d)} \partial_i ^2 \phi^2\right].
\label{Kd}
\end{equation}
Using the following identities:
\begin{equation}
(\partial_i \phi)^2=(\partial_r \phi)^2- \frac{\phi}{r^2}\Delta_{S^{d-2}}\phi,
\end{equation}
where we have partially integrated the angular piece, and
\begin{equation}
\partial_i^2 \phi^2 =\frac{1}{r^{d-2}}\partial_r \left(r^{d-2} \partial_r \phi^2\right)+\frac{1}{r^2}\Delta_{S^{d-2}}\phi^2,
\end{equation}
we arrive at
\begin{equation}
\begin{split}
K=\frac{1}{2}\sum_{\ell \vec{m}}\int dr r^{d-2} \left(\frac{R^2-r^2}{2 R}\right)&\left\{ \pi_{\ell \vec{m}}^2+ ( \partial_r \phi_{\ell \vec{m}})^2+\frac{\ell(\ell+d-3)}{r^2} \phi_{\ell \vec{m}} ^2- \right.\\
& \left. - \frac{(1-d/2)}{(1-d)} \left[\partial_r ^2  \phi_{\ell \vec{m}}^2+\frac{d-2}{r}\partial_r \phi_{\ell \vec{m}} ^2\right]\right\},
\end{split}
\end{equation}
In terms of the canonically conjugated operators,
\begin{equation}\label{Kred}
\begin{split}
K=\frac{1}{2}\sum_{\ell \vec{m}}\int dr  \left(\frac{R^2-r^2}{2 R}\right)&\left\{ \widetilde{\pi}_{\ell \vec{m}}^2+ (\partial_r \widetilde{\phi}_{\ell \vec{m}})^2+\frac{\mu_d(\ell)}{r^2} \widetilde{\phi}_{\ell \vec{m}}^2- \right.\\
& \left. -\frac{d-2}{2(d-1)}\left[3\partial_r\left(\frac{ \widetilde{\phi}_{\ell \vec{m}} ^2}{r}\right)+r\partial_r^2\left(\frac{ \widetilde{\phi}_{\ell\vec{m}} ^2}{r}\right)\right]\right\},
\end{split}
\end{equation}
Note that the second line of (\ref{Kred}), together with the prefactor $(R^2-r^2)$, is a total derivative in disguise. 
Hence,
\begin{equation}
\begin{split}
K=\frac{1}{2}\sum_{\ell \vec{m}}&\left\{\int_0^R dr  \left(\frac{R^2-r^2}{2 R}\right)\left[ \widetilde{\pi}_{\ell \vec{m}}^2+ (\partial_r \widetilde{\phi}_{\ell \vec{m}})^2+\frac{\mu_d(\ell)}{r^2} \widetilde{\phi}_{\ell \vec{m}} ^2\right]- \right.\\
& \left. -\frac{d-2}{2(d-1)}\left[\frac{(R^2-r^2)}{2R}r\partial_r\left(\frac{\widetilde{\phi}_{\ell\vec{m}}^2}{r}\right)+R \left(\frac{\widetilde{\phi}_{\ell\vec{m}}^2}{r}\right)\right]\biggr\rvert_0 ^{R}\right\},
\end{split}
\label{Kscalar}
\end{equation}
The boundary terms (coming from the improving) can be interpreted in general, as an ambiguity in the definition of modular Hamiltonian in a region,  and safely ignored as explained in \cite{Casini:2019qst}. Consequently, the modular Hamiltonian of the $d$ dimensional free scalar is
\begin{equation}
K=\sum_{\ell \vec{m}}K_{\ell\vec{m}},
\end{equation}
where
\begin{equation}\label{Klm}
K_{\ell\vec{m}}=\frac{1}{2}\int_0^R dr  \left(\frac{R^2-r^2}{2 R}\right)\left[ \widetilde{\pi}_{\ell \vec{m}}^2+ (\partial_r \widetilde{\phi}_{\ell \vec{m}})^2+\frac{\mu_d(\ell)}{r^2} \widetilde{\phi}_{\ell \vec{m}} ^2\right]
\end{equation}
can be interpreted as the modular Hamiltonian for the vacuum of (\ref{action1}) in a segment. This identification rests on the fact that the theory decomposes into independent sectors, labeled by the angular modes, so the state must write as the direct product of the states pertaining to each sector. But, most remarkably, this modular Hamiltonian is still local in the energy density. In other words, (\ref{Klm}) agrees with the general expression (\ref{KCFT}) in spite of the reduced one dimensional theories being non conformal. In the next section we analyse this in detail, paying attention to the symmetries which survive the dimensional reduction.

Provided that (\ref{Klm}) defines the reduced state of a free field theory, Wick's theorem guarantees it can be expressed in terms of the two-point correlators. In fact, for a Gaussian state with modular Hamiltonian
\begin{equation}
K=\int_V d^{d-1} x_1 d^{d-1}x_2 \left[\phi(x_1) M(x_1, x_2) \phi(x_2)+\pi(x_1) N(x_1, x_2) \pi(x_2)\right],
\end{equation}
and correlators
\begin{equation}
X=\langle \phi(x_1) \phi(x_2) \rangle\, , \quad P=\langle \pi(x_1) \pi(x_2) \rangle\, ,
\end{equation}
the following relation must be satisfied \cite{Casini:2009sr}\footnote{Here the product is a bi-local function constructed as $$ \left[M.X\right](x_1, x_2)\equiv \int_V d y M(x_1, y) X(y, x_2)$$}
\begin{equation}\label{MXPN}
M.X=P.N
\end{equation}
In the case at hand, 
\be
M(r,r')=-2\pi \delta(r-r')\left[\beta(r)\partial_r^2+\partial_r\beta(r) \partial_r-\beta(r) \frac{\mu}{r^2}\right]
\label{M1d}
\ee
and
\be 
N(r,r')=2\pi\delta(r-r')\beta(r)\,.
\label{N1d}
\ee
Meanwhile, the explicit form of the correlators for the one dimensional theory (\ref{action1}) is \cite{Saharian:2000mw}
\begin{equation}
X(r_1, r_2)=\frac{\Gamma\left[\ell+d/2-1\right]}{2\Gamma\left[\frac{1}{2}\right]\Gamma\left[\ell+\frac{d-1}{2}\right]}\left(\frac{r_1}{r_2}\right)^{\ell+\frac{d}{2}-1} {}_2 F_1 \left[ \frac{1}{2}, \ell+\frac{d}{2}-1; \ell+\frac{d-1}{2}; \left(\frac{r_1}{r_2}\right)^2\right]\, ,
\end{equation}
\begin{equation}
\begin{split}
P(r_1, r_2)&=\frac{2 \Gamma(\ell+d/2)}{\Gamma\left[\frac{1}{2}\right]\Gamma\left[\ell+\frac{d-1}{2}\right](r_2^2-r_1^2)}\left(\frac{r_1}{r_2}\right)^{\ell+\frac{d}{2}-1}\left(A ~ {}_2 F_1 \left[ \frac{1}{2}, \ell+\frac{d}{2}; \ell+\frac{d-1}{2}; \left(\frac{r_1}{r_2}\right)^2\right]\right.\\
& \left. + B ~ {}_2 F_1 \left[ -\frac{1}{2}, \ell+\frac{d}{2}; \ell+\frac{d-1}{2}; \left(\frac{r_1}{r_2}\right)^2\right]\right)\, ,
\end{split}
\end{equation}
where $A=\left(\ell+\frac{d-1}{2}\right)\left(1-r_1^2/r_2^2\right)-1$, $B=1-\ell-d/2$, and $r_1<r_2$. Using these concrete expressions it is possible to check that (\ref{MXPN}) indeed holds.
\section{Symmetries}\label{sec:symmetries}

The locality of (\ref{Klm}) suggests the existence of a symmetry with a conserved current such that the modular Hamiltonian is the corresponding Noether charge. This has to be an endomorphism in the causal wedge of the region, and must point in the time direction at $t=0$. For CFTs in spheres in any dimensions, this is the conformal transformation that maps the spherical boundary in itself. For an interval $(0,R)$ in the half line, whose causal wedge is a half diamond, the symmetry transformation leaves the boundary point $r=R$ fixed. The identification of this symmetry in the present case is the natural path to justify the locality of (\ref{Klm}). With this aim, we first discuss the symmetries of the reduced theories with action (\ref{action1}).

The symmetries of (\ref{action1}) are a subgroup of the conformal transformations inherited from higher dimensions, in particular those which involve only the time and radial coordinates, and that map the line $r=0$ into itself. These are
\begin{itemize}
\item Time translations: 
\begin{equation}\label{translation}
t\rightarrow t + t_0
\end{equation}

\item Dilatations:
\begin{equation}
(t,r)\rightarrow (\lambda t, \lambda r)
\end{equation}

\item Special conformal transformations with parameter $b^{\mu}= \frac{\alpha}{R}\hat{e}^{\mu}_t$:
\begin{equation}\label{STC}
(t,r)\rightarrow \left(\frac{t R^2+\alpha R(t^2-r^2)}{R^2+2\alpha R t+\alpha^2 (t^2-r^2)}, \frac{r R^2}{R^2+2\alpha R t+\alpha^2(t^2-r^2)}\right).
\end{equation}
Infinitesimally, that is, if we set $\alpha= \epsilon<<1$, then 
\begin{equation}
(t,r)\rightarrow  \left( t- \epsilon (t^2+r^2)/R,r-2\epsilon t r/R\right).
\end{equation}
\end{itemize}
The generators of the transformations listed above are $P_0=i\partial_t$, $D=i (t\partial_t+r\partial_r)$, and $K_0=i\left((t^2+r^2)\partial_t+2 t r\partial_r\right)$ respectively. These close an $sl(2, \mathbb{R})$ algebra, which can be expressed in a more suggestive way identifying $L_{-1}\equiv P_0$, $L_0\equiv D$, $L_1\equiv K_0$, so that

\begin{equation}
i\left[L_m,L_n\right]_{LB}=(m-n)L_{m+n}.
\end{equation}

Just for completeness, we note that one would have expected the original conformal group $SO(d,2)$ to break into $SO(2,2)\sim SL(2,\mathbb{R})\otimes \overline{SL(2,\mathbb{R})}$ \cite{Jackiw:2011vz,Chamon:2011xk}, with six generators. In fact, besides the three generators already mentioned, there are three more that do not mix the angular coordinates with $(t,r)$, associated to 
\begin{itemize}
\item Translations in the radial direction: 
\begin{equation}
r\rightarrow r+ r_0
\end{equation}

\item Boosts: 
\begin{equation}
(t,r)\rightarrow (t+ \epsilon r,r+\epsilon t)
\end{equation}

\item Special conformal transformations with parameter $b^{\mu}= \frac{\alpha}{R}\hat{e}^{\mu}_r$
\begin{equation}
(t,r)\rightarrow \left(\frac{t R^2}{R^2-2\alpha R r - \alpha^2 (t^2 -r^2)}, \frac{r R^2+\alpha R (t^2-r^2)}{R^2-2\alpha R r- \alpha^2 (t^2-r^2)}\right).
\end{equation}
\end{itemize}
These are $\hat{e}^{\mu}_r P_{\mu}$, $\hat{e}^{\mu}_r M_{0 \mu}$ and $\hat{e}^{\mu}_r K_{\mu}$, respectively. However, it is easy to see that they fail to become symmetries of the dimensionally reduced theory. 

Then, the modular symmetry of the reduced theories we are looking for must be a particular composition of the identified symmetry transformations (\ref{translation}) - (\ref{STC}).

On the other hand, we know that the modular symmetry for the parent conformal theory is associated to the generator of the boosts as seen from the domain of dependence of the ball \cite{VanRaamsdonk:2016exw},
\begin{equation}
\zeta=\frac{\pi}{R}\left[(R^2-t^2-\vert \vec{x}\vert^2)\partial_t-2 t x^i \partial_i\right]\,.
\end{equation}
In fact, comparing with (\ref{translation}) and (\ref{STC}), we notice that this transformation in the semi infinite line is the composition of a time translation of parameter $\epsilon\pi R$ and a special conformal transformation of parameter $\epsilon\frac{\pi}{R}$. Let us check this explicitly.

In spherical coordinates, the infinitesimal transformation reads
\begin{equation}\label{symmetry}
\begin{split}
& t\longrightarrow t'=t +\epsilon \frac{\pi}{R}(R^2-t^2-r^2)\\
& r\longrightarrow r'=r +\epsilon \frac{\pi}{R}(-2 t r)\\
& \Omega\longrightarrow \Omega'=\Omega
\end{split}
\end{equation}

Since the invariance of the kinetic term is guaranteed, we need only to check the invariance of the quadratic term $dt dr /r^2$, which is less evident. On the one hand, we have
\begin{eqnarray}
dt dr &=& dt' dr'\begin{vmatrix}
\frac{\partial t}{\partial t'} & \frac{\partial t}{\partial r'}\\
\frac{\partial r}{\partial t'} & \frac{\partial r}{\partial r'}
\end{vmatrix}= dt' dr' \begin{vmatrix}
1+2\pi\epsilon	t'/R +\mathcal{O}(\epsilon^2) & 2\pi\epsilon r'/R+\mathcal{O}(\epsilon^2)\\2\pi\epsilon r'/R+\mathcal{O}(\epsilon^2) & 1+2\pi\epsilon t'/R+\mathcal{O}(\epsilon^2)
\end{vmatrix} \nonumber\\
&\sim & dt' dr' (1+4\pi\epsilon t'/R).
\end{eqnarray}
On the other hand, 
\begin{equation}
\frac{1}{r^2}\sim \frac{1}{(r'+2\pi \epsilon t' r' /R)^2}= \frac{1}{r'^2}(1-4\pi\epsilon t'/R+\mathcal{O}(\epsilon^2)).
\end{equation}
Hence, 
\begin{equation}
\frac{dt dr}{r^2}=\frac{dt' dr'}{{r'} ^2}+\mathcal{O}(\epsilon^2).
\end{equation}
By Noether's theorem, there must exist a conserved current associated to (\ref{symmetry}), which is of the form\footnote{We remove the tildes and the angular mode labels to avoid cluttering.}
\begin{equation}
j^{\mu}=\left(\frac{\delta \mathcal{L}}{\delta (\partial_{\mu}\phi)}\partial_{\nu}\phi-\mathcal{L}\delta^{\mu}_{\nu}\right)\zeta^{\nu},
\end{equation}
or, in components,
\begin{equation}
j^t=\frac{1}{2}\left[(\partial_t\phi)^2+(\partial_r \phi)^2+\frac{\mu}{r^2}\phi^2\right]\frac{(R^2-t^2-r^2)}{R}-2 \frac{t r}{R} \partial_r\phi\partial_t\phi,
\end{equation}
\begin{equation}
j^r=\frac{1}{2}\left[(\partial_t\phi)^2+(\partial_r \phi)^2-\frac{\mu}{r^2}\phi^2\right]2\frac{tr}{R} -\partial_r\phi\partial_t\phi \frac{(R^2-t^2-r^2)}{R}.
\end{equation}

Finally, the current above corresponds to a modular Hamiltonian 
\begin{equation}
\begin{split}
K_{\ell \vec{m}} &=\int_0^R dr j_0 (t=0, r)\\
& = \frac{1}{2}\int_0^R dr  \left(\frac{R^2-r^2}{2 R}\right)\left[ \widetilde{\pi}_{\ell \vec{m}} ^2+ (\partial_r \widetilde{\phi}_{\ell \vec{m}})^2+\frac{\mu_d(\ell)}{r^2} \widetilde{\phi}_{\ell \vec{m}} ^2\right],
\end{split}
\end{equation}
the same as  (\ref{Klm}) deduced in the previous section from different arguments.

\section{Modular Hamiltonian and entropy}\label{sec:entropy}
In this section we study the spectrum of the modular Hamiltonian (\ref{Klm}). Solving the eigenfunction problem 	allows us to compute the entanglement entropy for an interval attached to the origin, as a function of the angular mode $\ell$ and the original spacetime dimension $d$. Then we sum over the modes and compare the result with the entanglement entropy of the $d$-sphere.
\subsection{Eigenfunctions}
In general, given a quadratic modular Hamiltonian of a region $V$,  of the form
\be
K=\int_V d^{d-1}x \,d^{d-1}x'\,\,\left(\phi(x)M(x,x')\phi(x')+\pi(x)N(x,x')\pi(x')\right),
\ee
with $M$ and $N$ real symmetric operators, the eigenfunctions are those of the right and left action of $M.N$, namely
\be
(N. M) u_{s}=s^2 u_{s}
\ee
\label{u}
\be
(M. N) v_{s}=s^2 v_{s}.
\label{v}
\ee
This leads to the alternative way of writing $K$
\begin{equation}
K=\int_V d^{d-1}x \int_0^{\infty} ds ~u_s(x)~ s~ v_s ^{*}(x).
\end{equation}
More concretely, the problem we are interested in is defined by (\ref{M1d}) and (\ref{N1d}), so the eigenfunctions $u$ and $v$ satisfy the following hypergeometric equations\footnote{For later convenience we renormalize the eigenvalues to absorb a factor $1/(2\pi)^2$.}
\be
\left[\beta^2\partial_r^2+\beta\partial_r\beta \partial_r-\beta^2 \frac{\mu}{r^2}\right]u_s= - s^2 u_s
\ee

\be
\left[\beta^2\partial_r^2+3\beta\partial_r\beta \partial_r+\left(\beta\partial_r^2\beta+(\partial_r\beta)^2-\beta^2 \frac{\mu}{r^2}\right)\right]v_s=- s^2 v_s
\ee
The solutions of these equations are\footnote{There is an additional independent solution, but we dismiss it because it does not go to zero at $r=0$, as mandated by the boundary conditions.}
\begin{equation}\label{eigenfunctions}
\begin{split}
u_s (r)&=N_u\left(\frac{r}{R}\right)^{-1+\frac{d}{2}+\ell}\left(\frac{R^2-r^2}{R^2}\right)^{-i s}{}_2F_1\left[\frac{1}{2}-i s, -1 + \frac{d}{2} + \ell - i s, \frac{d}{2}-\frac{1}{2} + \ell,\frac{r^2}{R^2}\right]\\
v_s(r)&=N_u\frac{R}{\beta(r)}u_s(r),
\end{split}
\end{equation}
where $N_u$ is a normalization constant.

Near $r\sim 0$ the solutions behave as,
\begin{equation}
u_s(r)\sim v_s(r)\propto r^{-1+\frac{d}{2}+\ell}
\end{equation}
in agreement with the classical profile (\ref{classical}), whereas near $r\sim R$ they behave as
\begin{equation}\label{nearR}
\begin{split}
u_s(r)&\sim N_u\left[\left( \frac{R-r}{R}\right)^{-i s}\alpha(s)+ c.c.\right]\\
v_s(r)&\sim N_u\left[\left( \frac{R-r}{R}\right)^{-1-i s}\alpha(s)+c.c.\right],
\end{split}
\end{equation}
with 
\begin{equation}
\alpha(s)=\frac{2^{-i s}\Gamma\left[\frac{d-1}{2}+\ell\right]\Gamma\left[2 i s\right]}{\Gamma\left[i s+\frac{1}{2}\right]\Gamma\left[i s+\ell+\frac{d}{2}-1\right]}.
\end{equation}
It is very important to keep in mind that there is a branch point at $r=R$. In fact, since the eigenfunctions must satisfy the orthogonality relation
\begin{equation}\label{orth}
\int_0^R dr u_s(r)v_{s'}^{*}(r)=\delta(s-s')\, ,
\end{equation}
in order to find out the normalization factor $N_u$ we substitute in (\ref{orth}) the leading terms in their Taylor series expansion (\ref{nearR}), because only the region near $r\sim R$ can contribute with a Dirac delta function. That results in
\begin{equation}\label{norm}
\int_0^R dr u_s(r)v_{s'}^{*}(r)\sim 2 \vert N_u\vert^2\text{Re}\left[ I(s-s') \alpha(s)\alpha^{*}(s')+ I(s+s') \alpha(s)\alpha(s')\right]\, ,
\end{equation}
where
\begin{equation}\label{Imenos}
\begin{split}
I(s)&\equiv\int_0^R dr \frac{R}{R-r}\exp{\left[-i s\log{\left(\frac{R-r}{R}\right)}\right]}\\
&=  R\left[\frac{i}{s}+\pi \delta(s)\right].
\end{split}
\end{equation}
Hence, neglecting the finite terms\footnote{We also neglect a contribution of the form $\delta(s+s')$, coming from the second term in (\ref{norm}), because it is non-zero only at $s=s'=0$.}, we have that (\ref{orth}) holds provided that
\begin{equation}
N_u=\frac{1}{\sqrt{2\pi R}\vert \alpha(s)\vert},
\end{equation}
save an overall phase that we set to one for convenience.
\subsection{The entropy}
As explained in \cite{Arias:2018tmw} in the context of the free chiral scalar, we can take advantage of the orthogonality relation to simplify the computation of the entanglement entropy, which can be expressed as a regularized integral over a small region behind the end point $r=R$, of the form
\begin{equation}\label{entropy}
\begin{split}
S(\ell,d)&=\int_0^{R-\epsilon} dr\,\int_0^{\infty} ds \,u_s(r)g(s)v_s^*(r)\\
&=-\,\lim\limits_{\delta s\rightarrow 0}\int_{R-\epsilon}^R  dr\,\int_0^{\infty} ds \,u_s(r)g(s)v_{s+\delta s}^*(r)\, ,
\end{split}
\end{equation}
with
\begin{equation}
g(s)= \frac{1+\coth(\pi s)}{2}\log\left( \frac{1+\coth(\pi s)}{2}\right)+ \frac{1-\coth(\pi s)}{2}\log\left( \frac{\coth(\pi s)-1}{2}\right)
\end{equation}
Note that since we expect the entanglement entropy of a QFT to diverge due to the short range correlations between modes at both sides of the boundary, we regularized it by introducing a small UV cutoff $\epsilon$. Furthermore, in going from the first to the second line of (\ref{entropy}) we shifted the $v$ sub index, summing over slightly off diagonal elements. For fixed $\delta s\neq 0$ the integral defined on the whole interval vanishes because of (\ref{orth}), leading to an integral just behind the boundary. This trick allows us to substitute the expansion (\ref{nearR}), which is much easier to integrate than the original solutions (\ref{eigenfunctions}). Finally, we get
\begin{equation}
S(\ell,d)=\frac{1}{6}\log{\frac{R}{\epsilon}}-\frac{1}{\pi}\int_0^{\infty} ds\,  g'(s) \text{Arg}(\alpha(s)),
\end{equation}
or, more explicitly,
\begin{equation}\label{s_l_d}
S(\ell,d)=\frac{1}{6}\log{\frac{R}{\epsilon}}-\frac{i \pi}{2}\int_0^{\infty} ds \frac{s}{\sinh^2(\pi s)}\log{\left(\frac{4^{ i s}\Gamma\left[i s\right]\Gamma\left[-1+d/2+\ell-i s\right]}{\Gamma\left[-i s\right]\Gamma\left[-1+d/2+\ell+i s\right]}\right)}
\end{equation}
The logarithmic coefficient $1/6$ is the expected result for a $(1+1)$ dimensional theory. Meanwhile, the constant term is expressed in terms of an integral that cannot be solved explicitly. For later convenience, we write it as a sum of two contributions, one that does not depend neither on the dimension nor on the angular mode
\begin{equation}
c\equiv-\frac{i \pi}{2}\int_0^{\infty} ds \frac{s}{\sinh^2(\pi s)}\log{\left(\frac{4^{ i s}\Gamma\left[i s\right]}{\Gamma\left[-i s\right]}\right)}\, ,
\end{equation}
and another which does depend on both parameters
\begin{equation}\label{f}
f(\ell,d)\equiv-\frac{i \pi}{2}\int_0^{\infty} ds \frac{s}{\sinh^2(\pi s)}\log{\left(\frac{\Gamma\left[-1+d/2+\ell-i s\right]}{\Gamma\left[-1+d/2+\ell+i s\right]}\right)}
\end{equation}
\begin{figure}[t]
\begin{center}  
\includegraphics[width=0.55\textwidth]{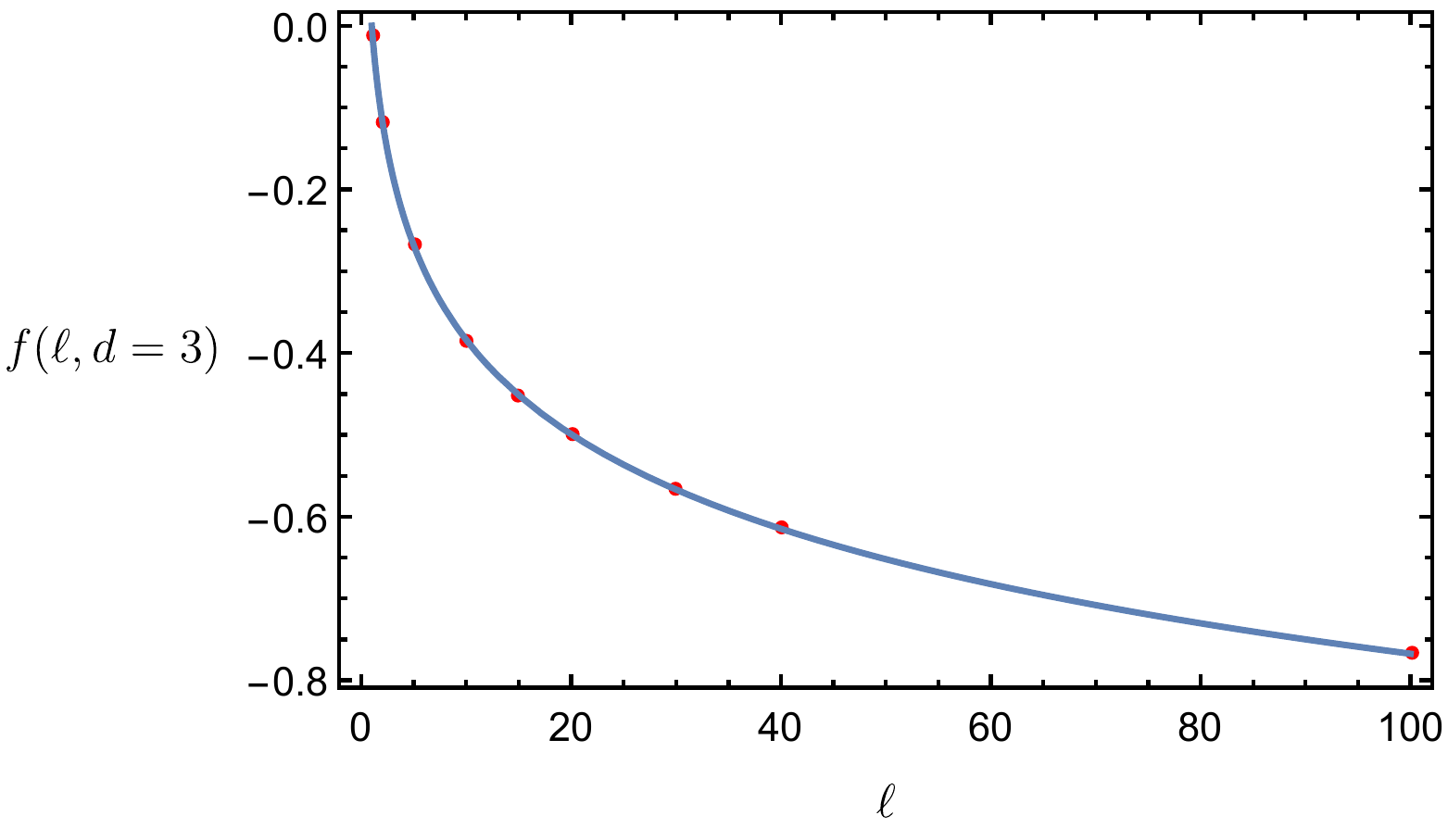}
\caption{Constant term of the entropy at $d=3$, as a function of the angular mode $\ell$. The red dots represent the exact numerical value of (\ref{f}), for $\ell=\lbrace 1, 2,5, 10, 15, 20, 30, 40, 100 \rbrace$. The blue curve corresponds to the fit $f(\ell,d=3)=c_0+ c_1\log{\ell}$, with $c_0=1.345\times 10^{-5}$ and $c_1=-0.1666$.}
\label{f_l_d3}
\end{center}  
\end{figure}
Although it is unfortunately impossible to find an analytic expression for the integral, for sufficiently large modes we can make use of the Stirling's approximation
\begin{equation}
\log{\Gamma(z)}\sim z \log{z}-z+\frac{1}{2}\log{\frac{2\pi}{z}}+\sum_{n=1}^{N-1}\frac{B_{2 n}}{2 n(2 n-1) z^{2 n-1}}, \quad \vert z\vert\rightarrow \infty
\end{equation}
to write 
\begin{equation}
\begin{split}
&\log{\left(\frac{\Gamma\left[-1+d/2+\ell-i s\right]}{\Gamma\left[-1+d/2+\ell+i s\right]}\right)}\sim -2 i s \log{\ell}+ \sum_{k=2}^{\infty} \sum_{m=1}^{\lfloor \frac{k+1}{2}\rfloor}\frac{(k-2)!}{\ell^{k-1}} a_{k,m}(s)\\
& +\frac{1}{2}\sum_{k=1}^{\infty} \sum_{m=1}^{\lfloor \frac{k+1}{2}\rfloor}\frac{(k-1)!}{\ell^{k}} a_{k,m}(s)+\sum_{n=1}^{\infty}\sum_{k=1}^{\infty} \sum_{m=1}^{\lfloor \frac{k+1}{2}\rfloor}\frac{B_{2 n} (2 n+k-2)!}{(2 n)!\ell^{2n+k-1}} a_{k,m}(s), \quad \ell>>1,
\end{split}
\end{equation}
where 
\begin{equation}
a_{k,m}(s)=  \frac{2 i(-1)^{k+m}}{(2m-1)! (k+1-2m)!}\left(-1+\frac{d}{2}\right)^{k+1-2 m}s^{2 m-1}.
\end{equation}
This means that the constant term grows logarithmically with the mode $\ell$, with corrections that decay as positive powers of $1/\ell$. In fact, performing the integration over the variable $s$ order by order in the expansion, we can straightforwardly check that the first few leading terms read
\begin{equation}
f(\ell,d)\sim -\frac{1}{6}\log{\ell}+ \frac{a_1}{\ell}+\frac{a_2}{\ell^2}+\frac{a_3}{\ell^3}+\frac{a_4}{\ell^4}+\mathcal{O}\left(\frac{1}{\ell^5}\right),
\end{equation}
with 
\begin{eqnarray}\label{a1}
a_1&=&\frac{1}{4}-\frac{d}{12}\\ \label{a2}
a_2&=&\frac{7}{40}-\frac{d}{8}+\frac{d^2}{48}\\ \label{a3}
a_3&=&\frac{3}{20}-\frac{7 d}{40}+\frac{d^2}{16}-\frac{d^3}{144}\\ \label{a4}
a_4&=&\frac{73}{560}-\frac{9 d}{40}+\frac{21 d^2}{160}-\frac{d^3}{32}+\frac{d^4}{384} 
\end{eqnarray}
Quite surprisingly, the logarithmic term already approximates $f(\ell,d)$ at $\ell\sim \mathcal{O}(1)$ very accurately, as shown in figure (\ref{f_l_d3}).

\begin{figure}[t]
\begin{center}  
\includegraphics[width=0.55\textwidth]{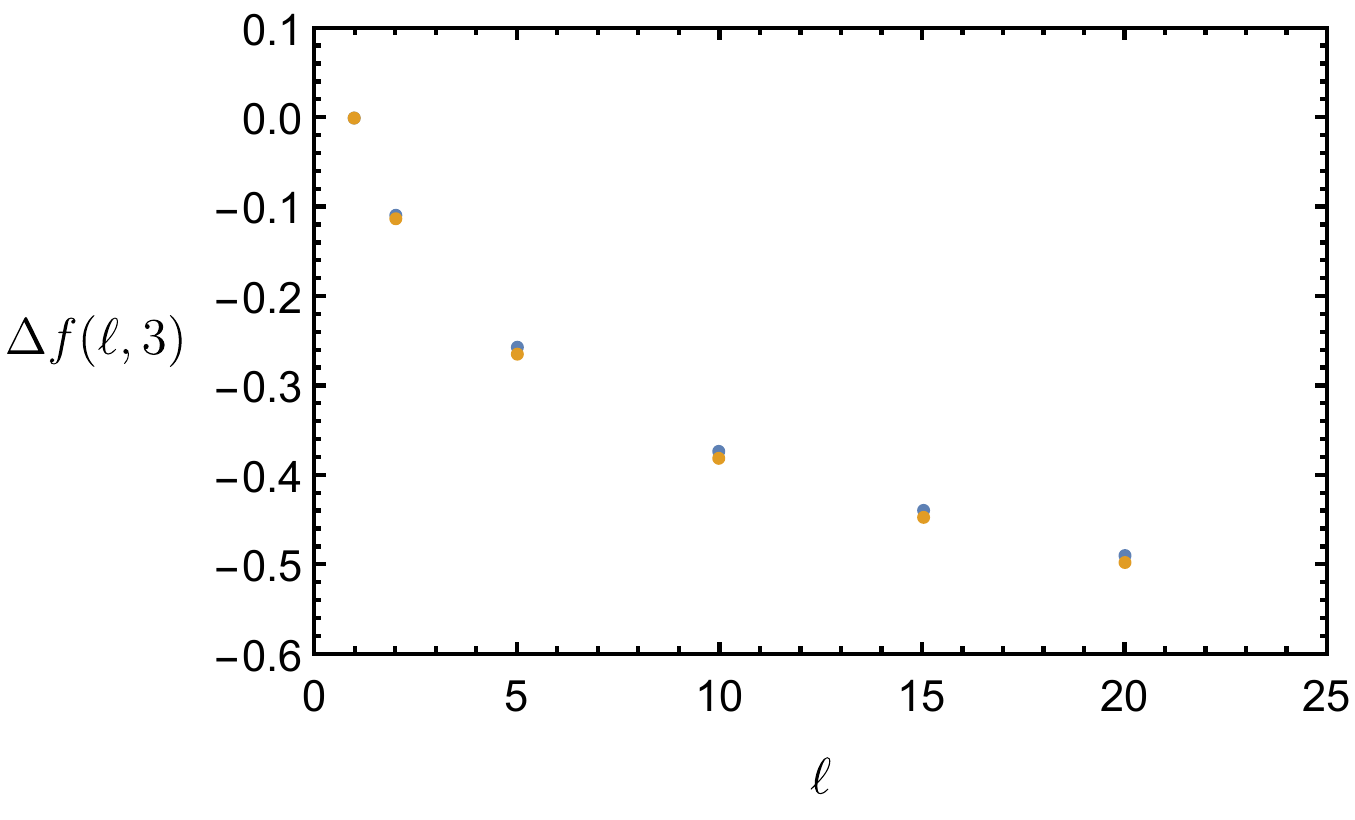}
\caption{$\Delta f(\ell,3)\equiv f(\ell,3)-f( \ell=1,3)$. Blue: direct numerical integration of (\ref{f}). Orange: calculation with a radial lattice regularization. $\ell=\lbrace 1, 2, 5, 10, 20\rbrace$}
\label{f_l_d3_lattice}
\end{center}  
\end{figure}

In figure (\ref{f_l_d3_lattice}) we compare the numerical value of (\ref{f}) with the one obtained from direct calculation in a radial lattice, again at $d=3$. Since the constant term depends on the regularization scheme, we subtract the one corresponding to $\ell=1$ and compare $\Delta f(\ell,3)\equiv f(\ell,3)-f( \ell=1,3)$. Although it is very hard to achieve good precision in the lattice\footnote{Roughly speaking, the value of $\ell$ gives a lower bound for the meaningful radios $R/\epsilon>>\ell$}, we find reasonable agreement. For example, for $\ell = 10$, numerical integration yields $\Delta f=-0.3737$, while the lattice computation gives $\Delta f=-0.3795$.
\subsection{Recovering the scalar entropy}
As discussed in section \ref{sec: mH}, the modular Hamiltonian of the free scalar in the sphere is equal to the sum over $\ell$ of the modular Hamiltonian pertaining to each one dimensional theory in the segment. Consequently, we expect that summing (\ref{s_l_d}) must necessarily reproduce the general structure for the entanglement entropy,
\begin{equation}\label{Sgen}
S=\begin{cases}
      \#\left(\frac{R}{\epsilon}\right)^{d-2}+ ... +c_{\text{log}} \log{\frac{R}{\epsilon}}, & d \quad\text{even} \\
       \#\left(\frac{R}{\epsilon}\right)^{d-2}+ ... + F, & d \quad \text{odd}
    \end{cases}
\end{equation}
that is, an infinite contribution controlled by the area term, and a universal piece, either in the form of a logarithmic coefficient in even dimensions, which is precisely the trace anomaly coefficient associated to the Euler density \cite{Cappelli:2000fe,Casini:2010kt,Casini:2011kv}, or a constant term in odd dimensions \cite{Klebanov:2011gs, Liu:2013una, Casini:2015woa}.

To show that this indeed holds, we need to introduce a cutoff in $\ell$ to regularize the sum. More concretely, we introduce a damping exponential so that
\begin{equation}
S=\sum_{\ell=0}^{\infty} \lambda(\ell, d) S(\ell, d) e^{-\ell \epsilon/R}.
\end{equation}
where
\begin{equation}
\lambda(\ell, d)=(2 \ell + d -3) \frac{(\ell + d -4)!}{\ell! (d-3)!}
\end{equation}
is the density of states. Note that this grows as $\ell^{d-3}$ for $\ell>>1$.

Given the complicated expression of the constant term $f(\ell,d)$, we approximate it by its large $\ell$ expansion, leading to
\begin{equation}\label{Ssum}
S= \sum_{\ell=1}^{\infty} \lambda(\ell, d) \left( \frac{1}{6}\log{\frac{R}{\epsilon}}+ c -\frac{1}{6}\log{\ell} +\sum_{j=1}^{j_{max}}\frac{a_j}{\ell^j}\right)e^{-\ell \epsilon/R} + \lambda(0, d) S(0, d) +\text{correction}.
\end{equation}
The correction above accounts for the error made when approximating $f(\ell,d)$ by its series expansion, truncated at $\mathcal{O}({\ell^{-j_{max}}})$.

It is straightforward to verify that (\ref{Ssum}) reproduces (\ref{Sgen}). For example, the divergent pieces come from terms with the general structure
\begin{equation}\label{divS1}
\sum_{\ell=1}^{\infty} \ell^p \left( \log{\frac{R}{\epsilon}}-\log{\ell}\right)e^{-\ell \epsilon/R}= -\Gamma'(p+1)\left( \frac{R}{\epsilon}\right)^{p+1}+\zeta(-p)\log{\frac{R}{\epsilon}}+\zeta'(-p),
\end{equation}
\begin{equation}\label{divS2}
\sum_{\ell=1}^{\infty} \ell^p e^{-\ell \epsilon/R}= p!\left( \frac{R}{\epsilon}\right)^{p+1}+\zeta(-p),
\end{equation}
with $\{p\mid p\in\mathbb{N}_0\wedge p\leq d-3\}$, and
\begin{equation}\label{divS3}
\sum_{\ell=1}^{\infty} \frac{1}{\ell}e^{-\ell \epsilon/R}= \log{\frac{R}{\epsilon}}.
\end{equation}
Note that the logarithmic term, only present in even dimensions, stems from (\ref{divS1}) and (\ref{divS3}). Based on this observation, it is worth pointing out that in order to compute the logarithmic coefficient we only need to take into account the first $d-1$ terms in the expansion of $f(\ell,d)$, that is, $j_{max}=d-2$. Subleading corrections give finite contributions at most.

\begin{figure}
\begin{center}  
\includegraphics[width=0.55\textwidth]{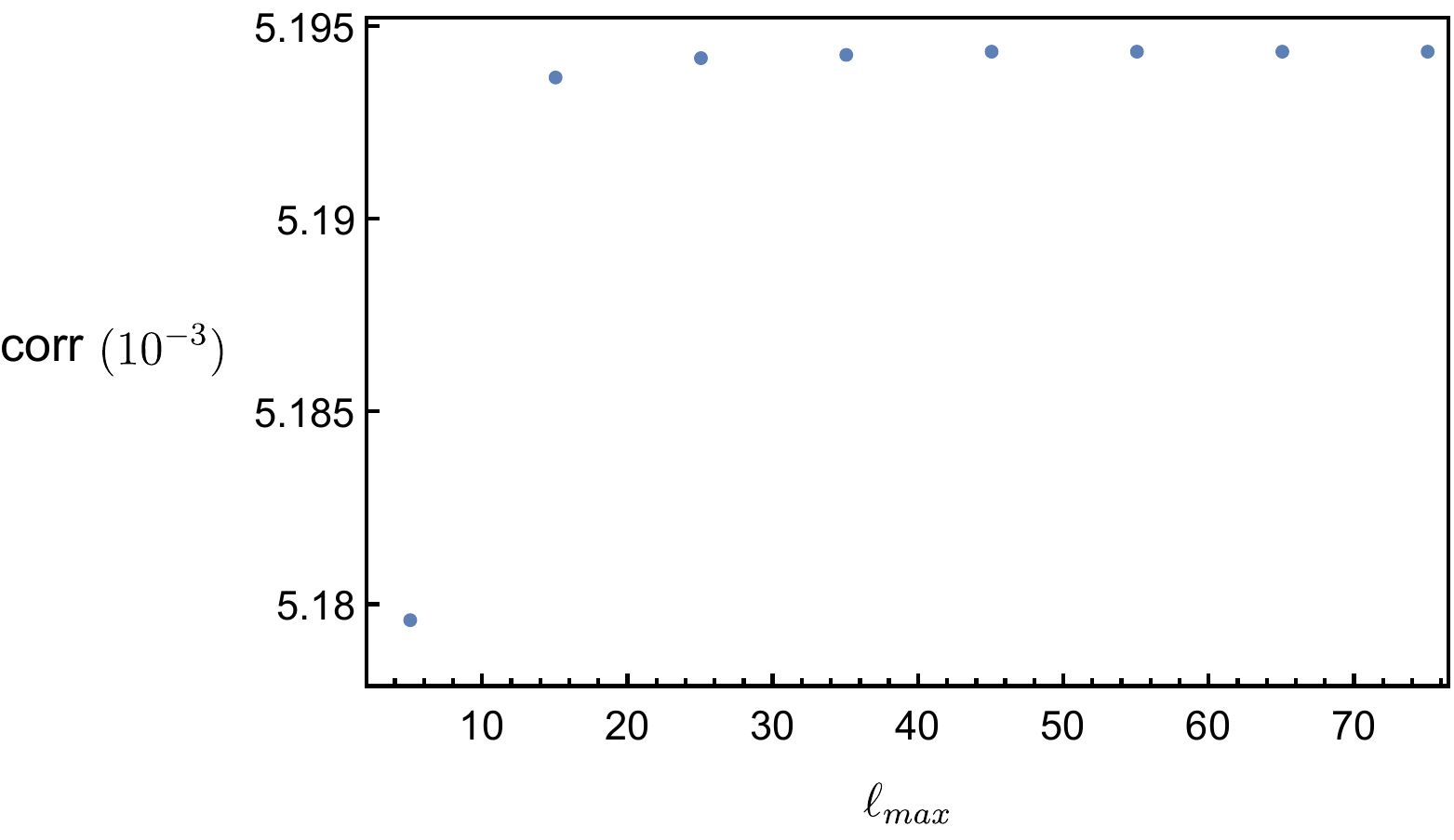}
\caption{Error made in the computation of the constant term $F$ when approximating $f(\ell,3)$ by $-\frac{1}{6}\log{\ell}+\frac{a_2}{\ell^2}$. $\ell_{max}$ is the greatest angular momentum that is summed over. We see that the correction converges very fast to $\sim 0.00519641$}
\label{corr}
\end{center}  
\end{figure}

Just to explicitly address some relevant specific cases, at $d=6$ we get
\begin{equation}
c_{\text{log}}(d=6)=\frac{29}{540} + a_1 + \frac{13}{6} a_2 + \frac{3}{2} a_3 + \frac{1}{3} a_4,
\end{equation}
and, substituting (\ref{a1}), (\ref{a2}), (\ref{a3}), (\ref{a4}), 
\begin{equation}
c_{\text{log}}(d=6)=\frac{1}{756},
\end{equation}
in agreement with the expected anomaly value. On the other hand, at $d=4$ we get
\begin{equation}
c_{\text{log}}(d=4)=\frac{1}{18} + a_1 + 2 a_2 ,
\end{equation}
which leads to the expected anomaly coefficient
\begin{equation}
c_{\text{log}}(d=4)=-\frac{1}{90}.
\end{equation}
The case of $d=3$ is different from the ones discussed above in that it has no logarithmic term and the universal piece in the entanglement entropy is associated to the constant term $F$. In fact, direct calculation yields, using (\ref{a1})
\begin{equation}
c_{\text{log}}(d=3)=2 a_1 =0.
\end{equation}
Regarding the constant $F$, the infinite tail in the $1/\ell$ expansion must in principle be taken into account. For that reason, we regularize the sum taking up to $j_{max}=2$ and then add a finite contribution which corrects the approximation, giving the exact value for the constant term in the series of $f(\ell,d)$. That is, 
\begin{equation}
\text{correction}=2\lim\limits_{\ell_{max}\rightarrow \infty}\sum_{\ell=1}^{\ell_{max}}\left(f(\ell, d=3)+\frac{1}{6}\log{\ell} - \frac{a_2}{\ell^2}\right)
\end{equation}
In figure (\ref{corr}) we plot the above correction as a function of $\ell_{max}$ and show that it converges very fast to
\begin{equation}
\text{correction}\sim 0.00519641
\end{equation}
According to (\ref{Ssum}), another term which contributes is 

\begin{equation}
f(0,3)=-\frac{i \pi}{2}\int_0^{\infty} ds \frac{s}{\sinh^2(\pi s)}\log{\left(\frac{\Gamma\left[1/2-i s\right]}{\Gamma\left[1/2+i s\right]}\right)}\sim 0.278435.
\end{equation}

Gathering all the pieces together, we finally get
\begin{equation}\label{F}
F= \frac{\pi^2}{3}a_2 +\frac{\zeta'(0)}{3} + f(0,3) + \text{correction}\sim -0.0638049,
\end{equation}
which is within $0.003 \%$ of the exact value \cite{Klebanov:2011gs,Casini:2015woa}. Note that the constant $c$ does not contribute at all to $F$.
\section{Final remarks}

In this article, we focused on theories in the semi infinite line constructed from the dimensional reduction of a free scalar in $d$ dimensions. Given that the decomposition of the parent theory $H$ into independent sectors $H_{\ell\vec{m}}$, labeled by the angular modes, also holds for the vacuum modular Hamiltonian in spheres $K=\sum_{\ell\vec{m}} K_{\ell\vec{m}}$, and provided that the vacuum state of the system is the product $\rho=\otimes\rho_{\ell\vec{m}}$, then, it is immediate to identify the modular Hamiltonian mode $K_{\ell\vec{m}}$ with the modular Hamiltonian of the one-dimensional reduced system $H_{\ell\vec{m}}$ in the interval $(0,R)$. Remarkably, the resulting modular Hamiltonian is local and proportional to the energy density, with the same weight function $\beta(r)=\frac{R^2-r^2}{2R}$ as the one characteristic of a CFT in a sphere $R$. 

We complemented the previous analysis with the study of the symmetries inherited from the $d$ dimensional conformal theory. This approach evidences the fact that the symmetry behind the locality of the reduced modular Hamiltonian is just the restriction to the semi infinite line of the original modular symmetry in $d$ dimensions.
We identified the conserved current associated to this symmetry transformation and checked that the $K_{\ell\vec{m}}$ found by dimensional reduction coincides with the Noether charge.

On the other hand, the spectral decomposition of the modular Hamiltonian leads to an analytic expression for the corresponding entanglement entropy (EE) which in turn, after summing over the angular modes, allowed us to recover the EE of the original $d$ dimensional theory in the sphere. 
To make sense of the sum, we used a novel regularization implemented by a damping exponential parametrized by the same cutoff $\epsilon$ that regularizes the radial coordinate. As we mentioned in the introduction, in a way, this procedure generalizes the one introduced by Srednicki in \cite{Srednicki:1993im} and provides an additional tool to calculate analytically the EE logarithmic coefficient in even dimensions, the universal constant term in $d=3$, among others.

It would certainly be interesting to explore in the future the modular Hamiltonian of non-conformal theories constructed from the dimensional reduction of other free theories, fermions for example. We expect that the decomposition of the parent theory into independent sectors must carry over unaltered, as well as the symmetry arguments which justify the resulting modular Hamiltonian is a conserved charge.

\section*{Acknowledgments}
We thank H.Casini, C.Fosco, E.Tonni and G.Torroba for discussions while this work was being carried out. This work was supported by CONICET, CNEA and Universidad Nacional de Cuyo, Instituto Balseiro, Argentina.




\bibliography{refs}

\bibliographystyle{utphys}

\end{document}